\documentclass[sigconf]{acmart}
\AtBeginDocument{%
  }

\usepackage{subfig}
\usepackage{tikz}
\usepackage{pgfplots}
\usepackage{pgf-pie}
\usepackage[most]{tcolorbox}
\usepgfplotslibrary{colorbrewer}

\pgfplotsset{compat = 1.15, cycle list/Set1-8} 
\pgfplotsset{
    /pgfplots/custom legend/.style={
    legend image code/.code={
        \draw[only marks,mark=square*] plot coordinates {(0.3cm,0cm)};
        },
    },
}
\usetikzlibrary{pgfplots.statistics, pgfplots.colorbrewer} 

\usepackage{pgfplotstable}

\copyrightyear{2025}
\acmYear{2025}
\setcopyright{cc}
\setcctype{by}
\acmConference[FSE Companion '25]{33rd ACM International Conference on the Foundations of Software Engineering}{June 23--28, 2025}{Trondheim, Norway}
\acmBooktitle{33rd ACM International Conference on the Foundations of Software Engineering (FSE Companion '25), June 23--28, 2025, Trondheim, Norway}\acmDOI{10.1145/3696630.3731674}
\acmISBN{979-8-4007-1276-0/2025/06}

\begin{document}

%%
%% The "title" command has an optional parameter,
%% allowing the author to define a "short title" to be used in page headers.
\title{An Empirical Study on the Impact of Gender Diversity on Code Quality in AI Systems}

% \author{Shamse Tasnim Cynthia \hspace{4mm} Banani Roy\\
% \normalsize Department of Computer Science, University of Saskatchewan, Canada\\
% \normalsize \{shamse.cynthia, banani.roy\}@usask.ca
% }

\author{Shamse Tasnim Cynthia}
\affiliation{%
  \institution{University of Saskatchewan}
  \state{Saskatchewan}
  \country{Canada}
}
\email{shamse.cynthia@usask.ca}

\author{Banani Roy}
\affiliation{%
  \institution{University of Saskatchewan}
  \state{Saskatchewan}
  \country{Canada}}
\email{banani.roy@usask.ca}

\begin{abstract}
  The rapid advancement of AI systems necessitates high-quality, sustainable code to ensure reliability and mitigate risks such as bias and technical debt. However, the underrepresentation of women in software engineering raises concerns about homogeneity in AI development. Studying gender diversity in AI systems is crucial, as diverse perspectives are essential for improving system robustness, reducing bias, and enhancing overall code quality. While prior research has demonstrated the benefits of diversity in general software teams, its specific impact on the code quality of AI systems remains unexplored. This study addresses this gap by examining how gender diversity within AI teams influences project popularity, code quality, and individual contributions. Our study makes three key contributions. First, we analyzed the relationship between team diversity and repository popularity, revealing that diverse AI repositories not only differ significantly from non-diverse ones but also achieve higher popularity and greater community engagement. Second, we explored the effect of diversity on the overall code quality of AI systems and found that diverse repositories tend to have superior code quality compared to non-diverse ones. Finally, our analysis of individual contributions revealed that although female contributors contribute to a smaller proportion of the total code, their contributions demonstrate consistently higher quality than those of their male counterparts. These findings highlight the need to remove barriers to female participation in AI development, as greater diversity can improve the overall quality of AI systems.
\end{abstract}

%%
%% The code below is generated by the tool at http://dl.acm.org/ccs.cfm.
%% Please copy and paste the code instead of the example below.
%%
\begin{CCSXML}
<ccs2012>
   <concept>
       <concept_id>10011007.10011074.10011134</concept_id>
       <concept_desc>Software and its engineering~Collaboration in software development</concept_desc>
       <concept_significance>500</concept_significance>
       </concept>
 </ccs2012>
\end{CCSXML}

\ccsdesc[500]{Software and its engineering~Collaboration in software development}

%%
%% Keywords. The author(s) should pick words that accurately describe
%% the work being presented. Separate the keywords with commas.
\keywords{diversity, AI systems, open-source, code quality}
%% A "teaser" image appears between the author and affiliation
%% information and the body of the document, and typically spans the
%% page.
% \begin{teaserfigure}
%   \includegraphics[width=\textwidth]{sampleteaser}
%   \caption{Seattle Mariners at Spring Training, 2010.}
%   \Description{Enjoying the baseball game from the third-base
%   seats. Ichiro Suzuki preparing to bat.}
%   \label{fig:teaser}
% \end{teaserfigure}

% \received{20 February 2007}
% \received[revised]{12 March 2009}
% \received[accepted]{5 June 2009}

%%
%% This command processes the author and affiliation and title
%% information and builds the first part of the formatted document.
\maketitle
% !--------------------------------------------------------!
\section{Introduction} \label{section:introduction}
% !--------------------------------------------------------!

The growing impact of Artificial Intelligence (AI) on industry and society is undeniable \cite{shams2023ai}. Generative AI, in particular, has reshaped how data is curated, analyzed, and used \cite{khan2024impact}, leading to an increase in open-source AI repositories on platforms like GitHub \cite{al2024enhancing}. 
As AI systems grow in complexity and societal influence, their reliability depends on the quality of their underlying code \cite{lenarduzzi2021software}. 
In this context, the composition of gender diverse development teams plays a critical factor in ensuring robust and sustainable system development \cite{adams2020diversity}. 
In software engineering, diverse teams have been shown to increase creativity, productivity, and performance \cite{scott2007power, tourani2017code, vasilescu2015gender, ostergaard2011does}.
However, the software engineering field faces a severe diversity crisis \cite{adams2020diversity}: less than 25\% of software engineers are women \cite{damian2024equity} which raises concerns about homogeneity in overall software development. This gap is alarming in AI systems, where high-quality code is essential for increasing productivity and efficiency for both developers and companies \cite{hansson2023code}. Moreover, maintaining the code quality in AI systems is not merely a technical concern but a sociotechnical imperative \cite{avgeriou2016managing, sculley2015hidden}, as poor-quality code can propagate biases, amplify technical debt, and undermine trust in AI outcomes \cite{lenarduzzi2021software}. Thus, understanding how gender diversity affects the code quality in AI systems is essential for ensuring both technical robustness and ethical development of these systems.

Several studies have examined the role of female contributors in software development, despite their persistent underrepresentation in collaborative environments \cite{zolduoarrati2021value, trinkenreich2022women, hoogendoorn2013impact, freire2021measuring, maheshwari2023review}. For example, Zolduoarrati et al. \cite{zolduoarrati2021value} found that female developers on Stack Overflow exhibit distinct knowledge-sharing behaviors, reflecting differences in orientation and attitudes toward technical engagement. 
% Similarly, Trinkenreich et al. \cite{trinkenreich2022women} highlighted the underrepresentation of women in open-source software (OSS), which contributes to career disadvantages and missed opportunities for skill development. Their review of 51 research articles, however, indicated that women participate in both code and non-code contributions within OSS projects. 
Trinkenreich et al.\cite{trinkenreich2022women} noted that while women are underrepresented in open-source software (OSS), they actively contribute to both code and non-code tasks.
Similarly, Terrell et al. \cite{terrell2017gender} found that women's pull requests are accepted at a higher rate than men's, suggesting a positive reception of their contributions. 
However, existing studies have not examined whether these contributions directly influence the quality of AI systems. While prior work highlights women’s involvement in collaborative software development, their impact on code quality within gender-diverse AI teams remains largely unexplored. Given the increasing reliance on AI systems, there is a critical need to investigate how gender diversity affects code quality at both the individual and team levels.

In this study, we conducted a comparative exploration of diverse and non-diverse AI systems, focusing on three key aspects: a) the impact of gender diversity on AI systems' popularity and community engagement, (b) differences in code quality between diverse and non-diverse AI systems, and (c) the quality of individual contributions within the diverse teams of AI systems. We addressed three research questions and presented three key findings. 

\textbf{RQ\textsubscript{1}} (\textit{diversity and impact}): \textbf{Does gender diversity impact the popularity and community engagement of AI systems?}  
We aim to determine whether AI teams with greater gender diversity achieve higher levels of popularity and community engagement compared to non-diverse teams. Understanding this impact is essential for assessing whether diversity enhances the visibility, adoption, and collaborative success of AI systems.

\textbf{RQ\textsubscript{2}} (\textit{quality comparison}): \textbf{How does the code quality in AI diverse teams compare to the code quality in AI non-diverse teams?}  
We seek to analyze whether gender-diverse AI teams produce higher-quality code compared to non-diverse teams. We aim to understand how team composition influences the maintainability and reliability of AI systems, which can provide insights into the potential benefits of diversity in the development of the systems.

\textbf{RQ\textsubscript{3}} (\textit{individual contribution quality}): \textbf{Within AI diverse teams, is the code authored by female developers of higher quality than the code authored by male developers?}  
We aim to assess the quality of individual contributions in multi-authored files by comparing code quality metrics between female and male contributors. This analysis seeks to uncover potential differences in coding practices and evaluate their implications for the overall quality and reliability of AI systems.

\noindent\textbf{Replication Package} is available in our online appendix \cite{replication_package}.
% \footnote{\url{https://anonymous.4open.science/r/HumanAISE2025-2636/}}.

% !--------------------------------------------------------------------!
\section{Related Work} \label{section:related_work}
% !--------------------------------------------------------------------!

Diversity in software engineering has been widely studied, with research consistently highlighting both its benefits and the persistent disparities in participation \cite{cavero2015evolution, bano2019gender, felizardo2021global, santana2021scientific, moldovan2024diversity, hosseini2021gender, rodriguez2021perceived}. 
The impact of diversity on software development productivity has been a key research focus. Vasilecu et al. \cite{vasilescu2015gender} found that gender and tenure diversity positively influence team productivity, as diverse teams integrate multiple perspectives to improve project outcomes. 
Similarly, Hoogendoorn et al. \cite{hoogendoorn2013impact} observed that gender-balanced teams outperformed male-dominated ones, reinforcing the value of inclusive teams.
However, despite these advantages, women remain significantly underrepresented in software engineering communities. Zolduoarrati et al. \cite{zolduoarrati2021value} investigated gender disparities in the Stack Overflow community, revealing sharp differences in participation, orientation, and knowledge-sharing behaviors between male and female contributors. 
Trinkenreich et al. \cite{trinkenreich2022women} further emphasized the limited role of women OSS, showing that less than 5\% of projects had women as core developers and that female-authored pull requests, while accepted at equal or higher rates than those of men, were disproportionately low in number. 
These findings align with Bosu et al. \cite{bosu2019diversity}, who identified a continued lack of gender diversity in popular OSS projects, particularly in leadership roles.
Some studies, such as Kazmi \cite{kazmi2014women} and Maheshwari \cite{maheshwari2023review} highlight socio-cultural barriers like work-life balance and impostor syndrome that hinder female participation. However, they suggest that mentorship and institutional support can help mitigate these barriers and improve women's retention and career progression.

While studies confirm the importance of diversity, some research has also explored gender biases in technical contributions and code review processes. 
For example, Catolino et al. \cite{catolino2019gender} examined the relationship between gender diversity and community smells, finding that teams with a balanced gender composition experienced fewer collaboration issues, indicating that diversity positively impacts team cohesion. 
Terrell et al. \cite{terrell2017gender} conducted a large-scale study on gender bias in OSS, demonstrating that women's pull requests were more frequently accepted than men's, but only when their gender was not publicly identifiable. Conversely, for contributors whose gender was visible, men had higher acceptance rates.
In the context of code reviews, Sultana et al. \cite{sultana2023code} found gender biases in 13 of 14 code review datasets, with women often facing longer delays before their contributions were accepted. 
Murphy-Hill et al. \cite{murphy2023systemic} found that women performed 25\% fewer code reviews than men and faced systemic bias in reviewer credential assignment and evaluation.
On the other hand, several studies have examined the challenges of maintaining high code quality in AI systems. 
Lenarduzzi et al. \cite{lenarduzzi2021software} identified key quality issues faced by AI developers, emphasizing that inadequate training leads to poor software practices. 
Yang et al. \cite{yang2023users} analyzed issues in open-source AI repositories, highlighting recurring problems that affect code maintainability. 
Similarly, Oort et al. \cite{van2021prevalence} and Zhang et al. \cite{zhang2022code} studied AI-specific code smells, identifying structural weaknesses that compromise long-term sustainability. These studies collectively underscore the need for rigorous development and maintenance practices in AI software engineering.

While previous research has explored gender diversity in software engineering, OSS, and AI research, the relationship between diversity and code quality in AI systems remains largely unexplored. Although studies like Terrell et al. \cite{terrell2017gender} and Trinkenreich et al. \cite{trinkenreich2022women} have analyzed gender disparities in contributions, they have not assessed whether gender-diverse teams produce higher-quality AI systems. Similarly, little is known about whether female-authored code differs in quality from male-authored code or how individual contributions vary between single-author and multi-author files. Additionally, the impact of diversity on repository popularity and community engagement in AI development remains an open question.
Our study addresses these gaps by conducting a comparative analysis of diverse and non-diverse AI repositories, evaluating their code quality, contribution patterns, and repository engagement. We also explore how gender influences individual contributions, particularly in multi-authored vs. single-authored files, offering new insights into the role of diversity in the development of AI systems.

% !-------------------------------------------!
\section{Methodology} \label{sec:methodology}
% !-------------------------------------------!
In this section, we describe the methodology of our study. The individual steps are as follows:

\begin{table}[t]
    \centering
    \small
    \caption{Analyzed features of the AI repositories \cite{fan2021makes}}
    \resizebox{0.8\linewidth}{!}{
        \begin{tabular}{|p{0.5cm}|p{1.75cm}|p{2cm}|p{3.5cm}|}
        \toprule
        ID      &                    Dimension           &   Feature name    &   Description  \\
       \midrule
        R1      &                Repository engagement   &   num-stars       &   Number of stars \\
        R2      &                                        &   num-forks       &   Number of forks \\
        R3      &                                        &   num-watchers    &   Number of watchers \\
        R4      &                                        &   num-open-issues  &   Number of open issues \\
        R5      &                                        &   num-closed-issues&   Number of closed issues \\
        R6      &                                        &   num-closed-prs   &   Number of closed pull requests \\
        R7      &                                        &   num-open-prs     &   Number of open pull requests \\   
        \midrule
        P1      &                        Code            &   num-code-files   &   Number of code files \\
        P2      &                                        &   num-code-lines   &   Number of lines of code \\
        P3      &                                        &   num-modules      &   Number of sub-directories in the root directory \\
        P4      &                        Reproducibility &   has-model       &   Whether the repository contains pre-trained models \\
        P5      &                                        &   has-shell       &   Whether the repository contains shell-scripts   \\
        P6      &                        Documentation   &   num-list        &   Number of lists in the README file  \\
        P7      &                                        &   num-code-blk    &   Number of code blocks in the README file    \\
        P8      &                                        &   num-inline-code &   Number of inline code elements in the README file   \\
        P9      &                                        &   num-img         &   Number of images in the README file \\
        P10      &                                       &   num-ghlink      &   Number of Github links in the README file   \\
        P11      &                                       &   has-license     &   Whether the repository has a license    \\
        \bottomrule
        \end{tabular}
    }
    \label{tab:features}
\end{table}
\vspace{-1em}
\subsection{Dataset Construction}
% ------------------------------------
% ------------------------------------
We examined women's involvement in AI systems using the dataset provided by Yang et al. \cite{yang2023users}. This dataset comprises 652 open-source AI repositories along with their corresponding papers sourced from \textit{PapersWithCode}\footnote{\url{https://paperswithcode.com/}}, a platform recognized for its reliability in archiving AI-related research, implementations, and datasets. According to the platform's assessment criteria, contributors marked the associated repositories as official \cite{yang2023users}. In our study, we focused on repositories hosted on GitHub that were finalized by the study's annotators, resulting in a subset of 576 repositories \cite{yang2023users}.

\subsection{Extracting the repository features} 
% -----------------------------------------------
% -----------------------------------------------

Next, we utilized the GitHub REST API\footnote{\url{https://docs.github.com/en/rest?apiVersion=2022-11-28}} to retrieve the recent repository features for the collected repositories. We successfully extracted seven features related to community engagement \cite{shameer2023relationship, al2020scoring} from 551 repositories, as the remaining repository addresses were no longer valid. These features included key repository attributes such as the number of contributors, stars, forks, watchers, and the total number of closed and open pull requests and issues. We referred to these features as \textit{repository-engagement} features in this study. Additionally, we extracted eleven features (e.g., number of code files, number of lines of code and number of modules) following the methodology of Fan et al. \cite{fan2021makes}, as these metrics are significantly associated with the popularity of AI repositories \cite{fan2021makes}. Therefore, we referred to these features as \textit{popularity} features. To assess diversity within repositories, we also collected contributor-specific metadata, including the number of contributions, usernames and locations. Table~\ref{tab:features} provides a detailed overview of our studied features.

\subsection{Identifying the genders of contributors}
% ---------------------------------------------------
% ---------------------------------------------------
Among the 551 repositories, we identified 228 repositories with more than one contributor. Since our study examines the impact of diversity on repository popularity and code quality, we excluded single-contributor repositories from our analysis.
To determine contributors' genders, we required their first and last names. However, as not all contributors provided their names, we followed the approach of Vasilescu et al. \cite{vasilescu2015gender} and included only repositories where at least 75\% of contributors’ genders could be identified. All selected repositories met this criterion.
For gender classification, we employed the genderComputer tool\footnote{\url{https://github.com/tue-mdse/genderComputer}}, which utilizes name and location data. Prior research on genderComputer reported a classification precision of up to 93\% \cite{vasilescu2015gender}. Using this tool, we identified the gender of 1,942 contributors, classifying 1,741 as male and 193 as female, with two contributors remaining unlabeled.
To evaluate the classification accuracy of the genderComputer tool, we manually verified a random sample of 500 labeled contributors by cross-referencing their GitHub and social media profiles. This validation identified seven misclassifications, which were subsequently corrected. The final dataset consisted of 1,745 male and 196 female contributors.

\subsection{Annotating team diversity} \label{subsec:annotate-diversity}
% ----------------------------------------------------------------------
% ----------------------------------------------------------------------
The Blau Index \cite{blau1977inequality} is a widely used metric for measuring diversity, particularly when multiple identity categories exist within a population \cite{vasilescu2015gender}. However, it assumes a more granular distribution of categories rather than a binary classification. Since our study focuses on distinguishing repositories based on the presence or absence of gender diversity, we adopted a binary classification approach: a repository is classified as \textbf{diverse} if it includes at least one female contributor, and a repository is classified as \textbf{non-diverse} if it consists exclusively of male or female contributors.

Our analysis identified 165 non-diverse repositories and 63 diverse repositories. Additionally, we observed that the total number of contributors in non-diverse teams ranged between 2 and 6 members. To ensure a fair comparison, we selected diverse repositories within the same contributor range for our calculations. As a result, our final dataset comprises 195 repositories (156 non-diverse and 39 diverse) for further analysis.
\vspace{-1em}
\begin{table}[htb]
    \centering
    \caption{Project category breakdown}
    \resizebox{0.5\linewidth}{!}{
        \begin{tabular}{|c|c|c|}
        \toprule
                        &  Category       &   \# of Repositories\\
        \midrule
        non-diverse     &   Small         &   80 \\
                        &  Medium         &   27 \\
                        &   Large         &   18 \\
        \midrule
        Diverse         &   Small         &   23 \\
                        &   Medium        &   5 \\
                        &   Large         &   8   \\
        \bottomrule
        \end{tabular}
    }
    \label{tab:project_category}
\end{table}
\vspace{-0.5em}

\vspace{-1em}
\begin{figure*}
    \centering
    \includegraphics[width=4in]{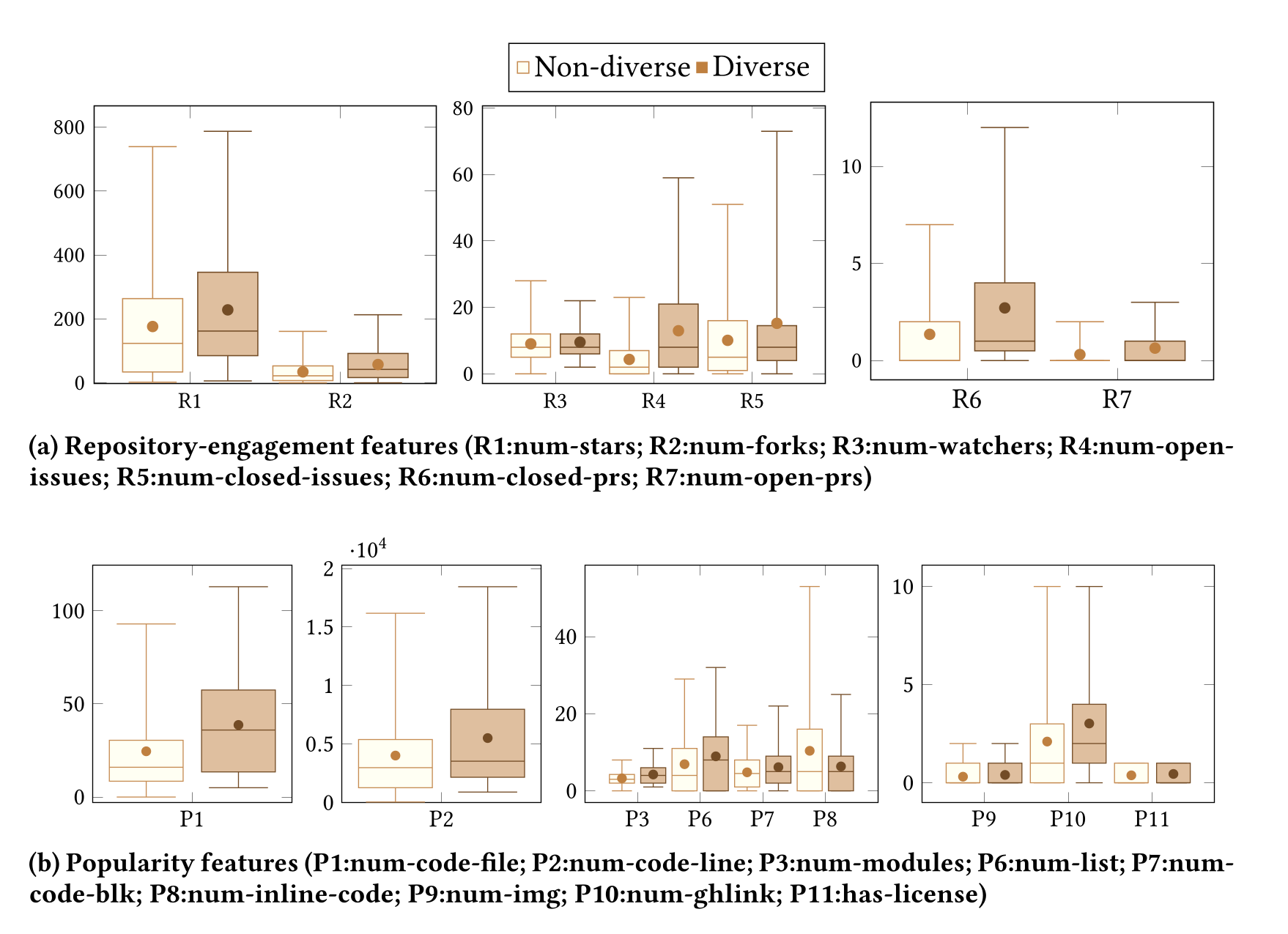}
    \caption{Repository-engagement and popularity features comparison of diverse and non-diverse AI repositories ($\bullet$ represents the mean value)}
    \label{fig:metric-comparison}
\end{figure*}
\vspace{-1em}

\subsection{Analyzing the impact of diversity in AI repositories}
% ----------------------------------------------------------------
% ----------------------------------------------------------------
We adopted the Wilcoxon rank-sum \cite{wilcoxon1992individual} test to compute the significance of the differences between the feature values of diverse and non-diverse groups. We also estimated the Cliff's $\delta$ \cite{cliff2014ordinal} of the difference between the diverse and non-diverse groups. According to the guideline for using the effect size, when $|\delta|$ is less than 0.147, between 0.147 and 0.33, between 0.33 and 0.474, and larger than 0.474, the effect size is considered negligible, small, medium, and large, respectively.

\subsection{Assessing the code quality of AI repositories} \label{subsec:RQ2}
% ----------------------------------------------------------------------------
% ----------------------------------------------------------------------------
We utilized SonarQube to evaluate the code quality of both diverse and non-diverse repositories. Among the widely used Automated Static Analysis Tools (ASAT) in open-source development (e.g., FindBugs and Checkstyle), SonarQube is the only tool that provides a Technical Debt Index \cite{vassallo2018context, lenarduzzi2020survey}. It assesses multiple code quality metrics, including code complexity, the presence of code smells, and compliance with predefined coding rules across various programming languages \cite{lenarduzzi2020some}. For our analysis, we selected (a) security hotspots, (b) number of code smells, (c) comment lines density (\% comments), (d) cyclomatic complexity and (e) cognitive complexity, to compare the code quality of diverse and non-diverse repositories.
These metrics are commonly used in research on source code quality \cite{nunez2017source} and are also measured by SonarQube. Notably, Code Smell is a composite metric encompassing various quality indicators, while Cognitive Complexity is a recently introduced metric designed to complement Cyclomatic Complexity by emphasizing code understandability \cite{campbell2018cognitive}.

For a structured comparison of code quality across different project scales, we categorized the AI repositories based on the number of lines of code (LOC) \cite{nunez2017source}. Specifically, repositories with a $<=5000$ LOC are categorized as small, those with $>5000$ and $<=10000$ LOC as medium and repositories with more than 10000 LOC are considered as large repositories. Since the majority of AI repositories are developed in Python, we excluded repositories in other programming languages. Table~\ref{tab:project_category} presents the final breakdown of repositories included in our analysis.

\subsection{Comparing code quality between male and female contributors}
% -----------------------------------------------------------------------
% -----------------------------------------------------------------------
First, we utilized git-blame\footnote{\url{https://git-scm.com/docs/git-blame}}, a Git feature, to identify the authors of each code file. For single-authored files, we categorized them as either male-authored or female-authored to facilitate a direct comparison of code quality. Next, we analyzed multi-authored files, separating male-authored and female-authored code segments to examine potential differences in quality metrics. In addition to the metrics mentioned in Section~\ref{subsec:RQ2}, we analyzed the Lines of Code (LOC) contributed by each author within these multi-authored files to assess individual contributions in a collaborative development setting. This analysis allowed us to examine potential differences in coding contributions between male and female developers within shared codebases.

% !------------------------------------------------!
\section{Study Finding} \label{sec:study_finding}
% !------------------------------------------------!
\subsection{Impact of Gender Diversity in AI Repositories (RQ\textsubscript{1})} \label{sec:RQ1}
% !------------------------------------------------------------------------------------------------!

\begin{figure*}
    \centering
    \includegraphics[width=4.8in]{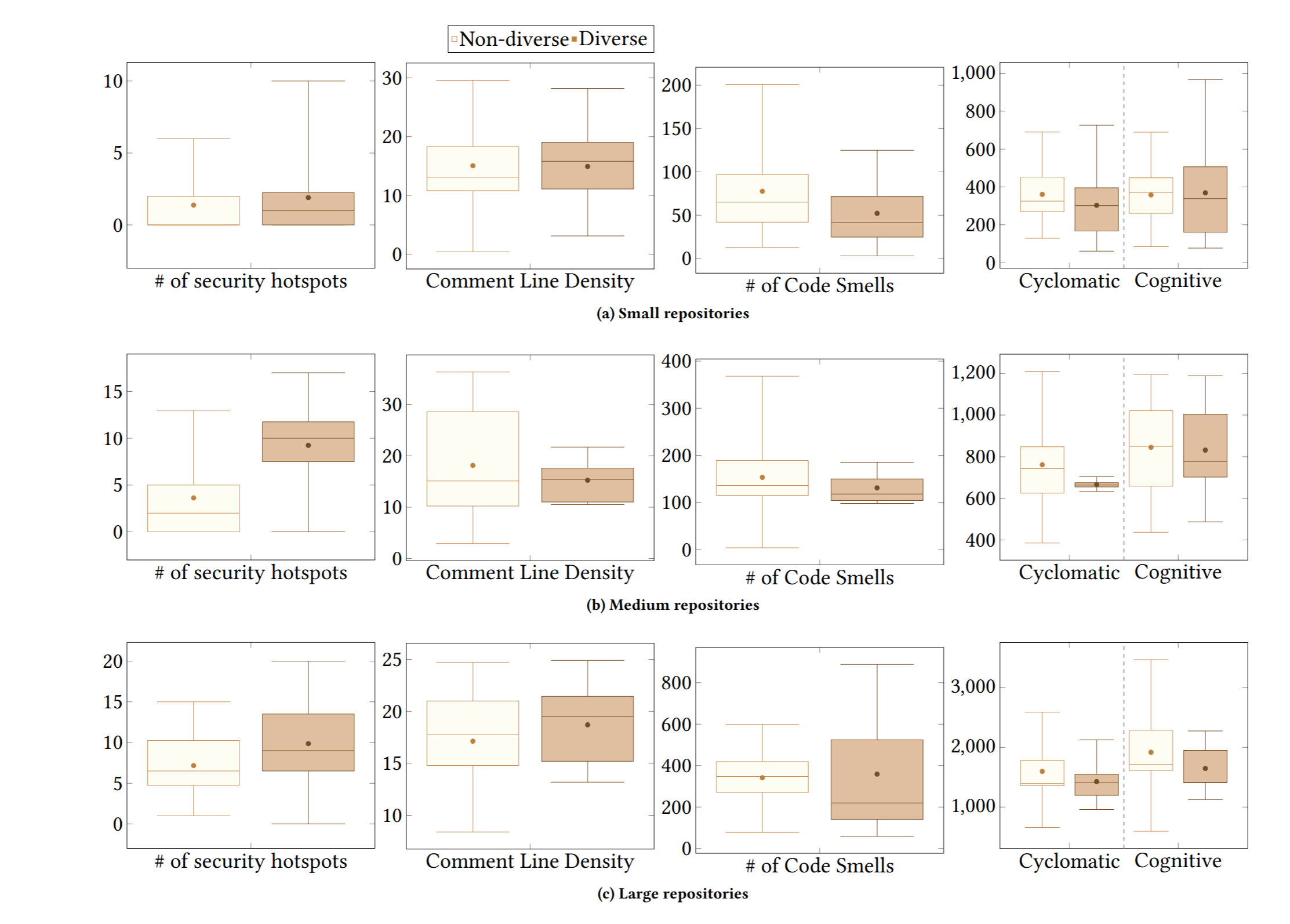}
    \caption{Code quality comparison of diverse and non-diverse AI repositories ($\bullet$ represents the mean value).}
    \label{fig:code-quality}
\end{figure*}

\vspace{-0.5em}
We examined the extracted features of both diverse and non-diverse AI repositories to analyze their impact on the repositories. Figure ~\ref{fig:metric-comparison} highlights notable patterns observed in diverse repositories. 
For repository-engagement features, we found that, with the exception of the \textit{num-watchers} feature, diverse repositories generally scored higher. For instance, in at least half of the AI repositories, the \textit{num-forks} feature in diverse AI repositories is almost two times higher than in non-diverse AI repositories, i.e., the median value of forks is 22 in non-diverse repositories and the median value of forks is 42 in diverse repositories. However, we observed little difference in the number of open and closed pull requests between the groups. Conversely, there are significant differences in the number of open and closed issues between the repositories.
On the other hand, for quartile analysis of the popularity features, we excluded the \textit{has-model} and \textit{has-shell} features from Fig.~\ref{fig:metric-comparison}, as their presence in both groups was minimal. Specifically, 153 out of 156 non-diverse repositories contained no pre-trained models, and 121 out of 156 lacked shell scripts. Similarly, in the diverse group, 35 out of 39 repositories had no pre-trained models, and 31 out of 39 contained no shell scripts. 
These findings show that our selected diverse and non-diverse repositories rarely included pretrained models and shell scripts. The few instances where these features were present suggest that they were handled by a small number of contributors, likely outliers in our dataset.
Examining the remaining popularity features from Fig.~\ref{fig:metric-comparison}, we observed that diverse AI teams tend to maintain larger repositories. For example, in at least half of the repositories, diverse teams maintain nearly twice as many code files (median: 26) as non-diverse teams (median: 16). In addition, both the \textit{num-code-line} and the \textit{num-modules} features are higher in diverse repositories. In terms of documentation, README files in diverse repositories include significantly more structured elements, such as lists, code blocks, and GitHub links, with a median of 8 lists in the README file compared to 4 in non-diverse repositories. This suggests that diverse AI teams prioritize clearer documentation and code readability, which may enhance maintainability and accessibility for contributors. Notably, there is little difference in the \textit{num-img} and \textit{has-license} features between the two repositories.

We applied the Mann-Whitney-Wilcoxon statistical test to assess the differences between diverse and non-diverse repositories. We also used Cliff's $\delta$ to quantify the effect sizes. Interestingly, significant differences were found for most of the repository-engagement features except the \textit{num-watchers}. While the effect sizes for these differences are generally small, the number of open issues shows a medium effect size. These findings imply that diverse repositories attract higher community engagement, as indicated by higher numbers of stars and forks. On the other hand, for the popularity features, the \textit{num-files}, \textit{num-code-lines}, \textit{num-modules}, and \textit{num-ghlink} showed statistically significant differences (p-values < 0.05) between diverse and non-diverse repositories. Our quartile analysis indicates that these differences have small to medium effect sizes. Although some effect sizes are small, the consistent pattern across multiple metrics suggests that diverse teams produce repositories with more detailed and modularized code.

These findings indicate that AI repositories developed by diverse teams tend to exhibit more extensive and structured characteristics compared to those non-diverse teams. Diverse repositories having statistically significantly higher values in repository-engagement features suggest that they might be receiving more attention or engagement from the community.
\vspace{-1em}
\begin{tcolorbox}[enhanced,attach boxed title to top center={yshift=-3mm,yshifttext=-1mm},
  colback=brown!60!black!2!white,colframe=brown!60!black!40!gray,colbacktitle=pink!20!white,coltitle=black,
  title=Summary of RQ1,fonttitle=\bfseries,
  boxed title style={size=small} ]
  Diverse AI repositories are statistically significantly different from non-diverse repositories with a non-negligible effect size in ten out of sixteen numeric features. Generally, diverse AI repositories have more code files, lines of code, and modules than non-diverse repositories. Furthermore, a higher number of repository-engagement features such as stars, forks, and issues, indicates that diverse AI repositories have more engagement from the community.
\end{tcolorbox}

% !--------------------------------------------------------------------------------------------------------!
\subsection{Code Quality in Diverse vs. Non-Diverse AI Repositories (RQ\textsubscript{2})} \label{sec:RQ2}
% !--------------------------------------------------------------------------------------------------------!

In the previous research question, our findings suggested that non-diverse teams may reveal less detailed code organization and documentation, potentially impacting repository popularity. Additionally, community engagement appeared lower in non-diverse repositories compared to diverse ones, which could influence code maintainability, collaboration efficiency, and overall repository quality.
Therefore, in this section, we employed SonarQube to analyze the code quality of both diverse and non-diverse repositories. This analysis aims to assess the impact of diversity on overall code quality by examining key software metrics.

% non-diverse repositories tend to have more bugs than diverse repositories. The gap is more prominent in small and medium repositories. Similarly,

As shown in Fig.~\ref{fig:code-quality}, we found significant differences in code quality between diverse and non-diverse repositories across all three categories (Section~\ref{subsec:annotate-diversity}). Specifically, non-diverse repositories showed a higher number of code smells, particularly in small and medium repositories. However, in large repositories, the distributions overlap more for diverse repositories. Additionally, both cyclomatic complexity and cognitive complexity were generally lower in diverse repositories. These findings suggest that gender-diverse teams incorporate a broader range of cognitive styles and problem-solving approaches, which leads to improved code quality \cite{page2008difference, park2015effects}. Furthermore, gender diversity may enhance communication styles and collaborative strategies, contributing to better software development practices. Catolino et al. \cite{catolino2019gender} also found that the presence of women reduces community smells, which in turn mitigates code smells and enhances overall software quality \cite{palomba2018beyond}.

Conversely, diverse AI repositories tend to have a higher number of security hotspots. However, since most security hotspots are precautionary warnings unless exploited \cite{yu2023towards}, this trend may originate from diverse teams triggering more SonarQube alerts due to variations in coding practices \cite{lix2022aligning, beckwith2005effectiveness} rather than actual security risks. Meanwhile, comment line density remains fairly consistent across both groups. This suggests that gender diversity does not have a significant influence on the documentation habits within the codebase.
Although we conducted a statistical test in \textbf{RQ1}, we did not perform one for this research question due to the limited sample size within categories, which could affect the reliability and statistical power of the results \cite{SO_question, button2013power}. 
\vspace{-1em}
\begin{tcolorbox}[enhanced,attach boxed title to top center={yshift=-3mm,yshifttext=-1mm},
  colback=brown!60!black!2!white,colframe=brown!60!black!40!gray,colbacktitle=pink!20!white,coltitle=black,
  title=Summary of RQ2,fonttitle=\bfseries,
  boxed title style={size=small} ]
    \textbf{Summary of RQ2:} Diverse AI repositories tend to show better code quality, specifically, fewer code smells and lower cyclomatic and cognitive complexity. However, due to limited sample sizes, these findings should be considered exploratory. Notably, diverse repositories also exhibit more security hotspots, while comment line ratios remain similar across both groups.
\end{tcolorbox}

% !--------------------------------------------------------------------------------------------------------------------------!
\subsection{Quality of Code Authored by Female vs. Male Developers in AI Diverse Teams (RQ\textsubscript{3})} \label{sec:RQ3}
% !--------------------------------------------------------------------------------------------------------------------------!

\begin{figure}
    \centering
    \includegraphics[width=2.5in]{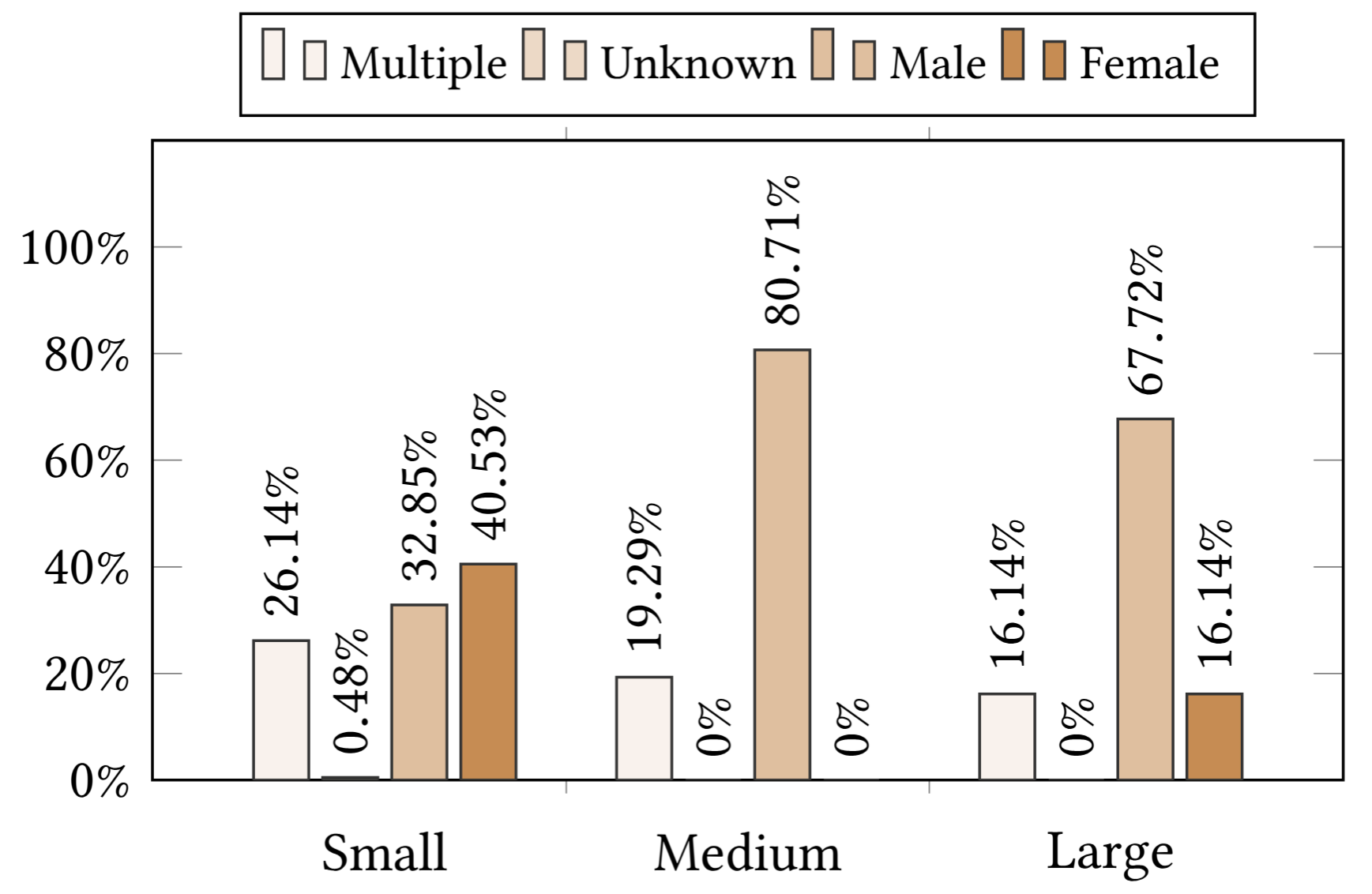}
    \caption{Ratio of multiple-authored, unknown-authored, male-authored and female-authored code files.}
\label{file_ratio}
\end{figure}

While our analysis of RQ2 demonstrates that gender-diverse teams generally produce higher-quality code than non-diverse teams, the individual contributions within these diverse teams remain unclear. Therefore, in this research question, we investigated whether the code authored by female contributors demonstrates higher quality than that authored by male developers within diverse AI repositories.
As shown in Fig.~\ref{file_ratio}, the male-authored code files are significantly higher in both medium and large repositories, whereas female-authored code files are comparatively less. Notably, in medium repositories, we observed no code files authored exclusively by female contributors. Similarly, in large repositories, only 16.14\% (169 out of 417) of files were authored solely by female contributors. However, small repositories exhibited the highest proportion of multiple-authored code files, suggesting a greater tendency for collaboration in these projects.   
Next, we examined the code quality of male-authored and female-authored files. Since no exclusively female-authored code files were found in medium-sized repositories, we excluded them from the comparative analysis. 

\begin{table}[h]
    \centering
    \caption{Quality metrics of male-authored and female-authored files in diverse repositories}
    \resizebox{0.7\linewidth}{!}{
        \begin{tabular}{p{3cm}p{1cm}p{1cm}p{1cm}p{1cm}}
            \toprule
            & \textbf{Small (M)} & \textbf{Small (F)} & \textbf{Large (M)} & \textbf{Large (F)} \\
            \midrule
            \textbf{Security Hotspots} & 144 & 12 & 100 & 5 \\
            \textbf{Comment (\%)} & 17.6 & 15.9 & 18.9 & 19.3 \\
            \textbf{Code Smells} & 533 & 308 & 2053 & 633 \\
            \textbf{Cyclomatic} & 1842 & 2548 & 9221 & 2401 \\
            \textbf{Cognitive} & 2077 & 2659 & 11022 & 2613 \\
            \bottomrule
        \end{tabular}
    }
    \label{tab:security_analysis}
\end{table}
As shown in Table~\ref{tab:security_analysis}, female-authored code appears to be less prone to SonarQube's security alerts. To further assess whether these security hotspots indicate actual vulnerabilities or conservative coding practices, we analyzed their severity. Our findings indicate that male-authored code files contain more security hotspots of medium priority than female-authored code.
Similarly, we found that female-authored code contains fewer code smells than male-authored code, which indicates more maintainable code structures. 
\vspace{-0.7em}

\vspace{-0.5em}
However, female-authored code in small repositories is more complex and more cognitively complex than that in large ones. A possible explanation is that in small repositories, female-authored code might incorporate more complex or experimental solutions, leading to higher cyclomatic and cognitive complexity. In contrast, in large repositories, where development processes, code reviews, and design standards tend to be more rigorous, female developers may adhere more closely to best practices, producing simpler, more modular code \cite{frluckaj2022gender}.
On the other hand, differences in commenting practices between male- and female-authored code files were minor across all cases.

\begin{figure}
    \centering
    \includegraphics[width=\linewidth]{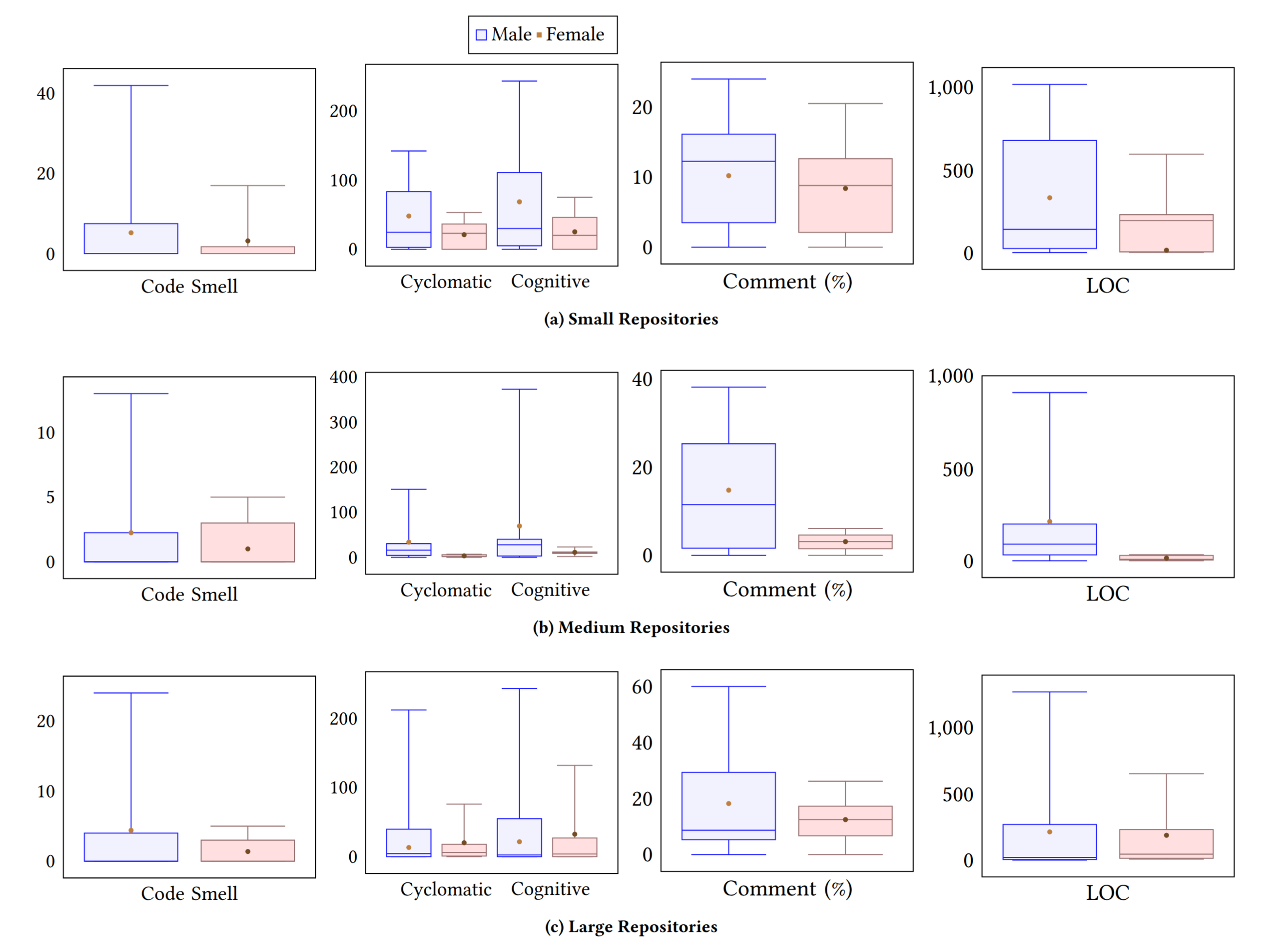}
    \caption{Code quality comparison of male-written and female-written code in multi-authored files.}
    \label{fig:multi-authored}
\end{figure}

Additionally, we analyzed the quality metrics of multi-authored files where both male and female developers contributed, as illustrated in Fig.~\ref{fig:multi-authored}. We excluded security hotspots due to zero values after outlier removal and incorporated the number of lines of code (LOC) to assess the ratio of female contributors' participation. The results indicate that female developers' contributions, although lower in volume (evidenced by fewer lines of code across all repositories), exhibit higher code quality. Specifically, code authored by female contributors demonstrates lower cyclomatic and cognitive complexity, fewer code smells, and a reduced code-comment ratio compared to their male counterparts. These findings suggest that female developers maintain higher coding standards even in collaborative environments.

\vspace{-0.5em}

\begin{tcolorbox}[enhanced,attach boxed title to top center={yshift=-3mm,yshifttext=-1mm},
  colback=brown!60!black!2!white,colframe=brown!60!black!40!gray,colbacktitle=pink!20!white,coltitle=black,
  title=Summary of RQ3,fonttitle=\bfseries,
  boxed title style={size=small} ]
  The analysis revealed that female-authored code files, though fewer in number, demonstrate fewer security hotspots, lower code smells, and reduced complexity in large repositories, indicating better maintainability. However, in small repositories, female-authored code shows higher complexity, potentially due to more experimental solutions. Moreover, in multi-authored files, female contributors show better coding standards despite contributing less.
\end{tcolorbox}

% !----------------------!
\section{Key Findings}
% !----------------------!

\begin{figure}
    \centering
    \includegraphics[width=0.7\linewidth]{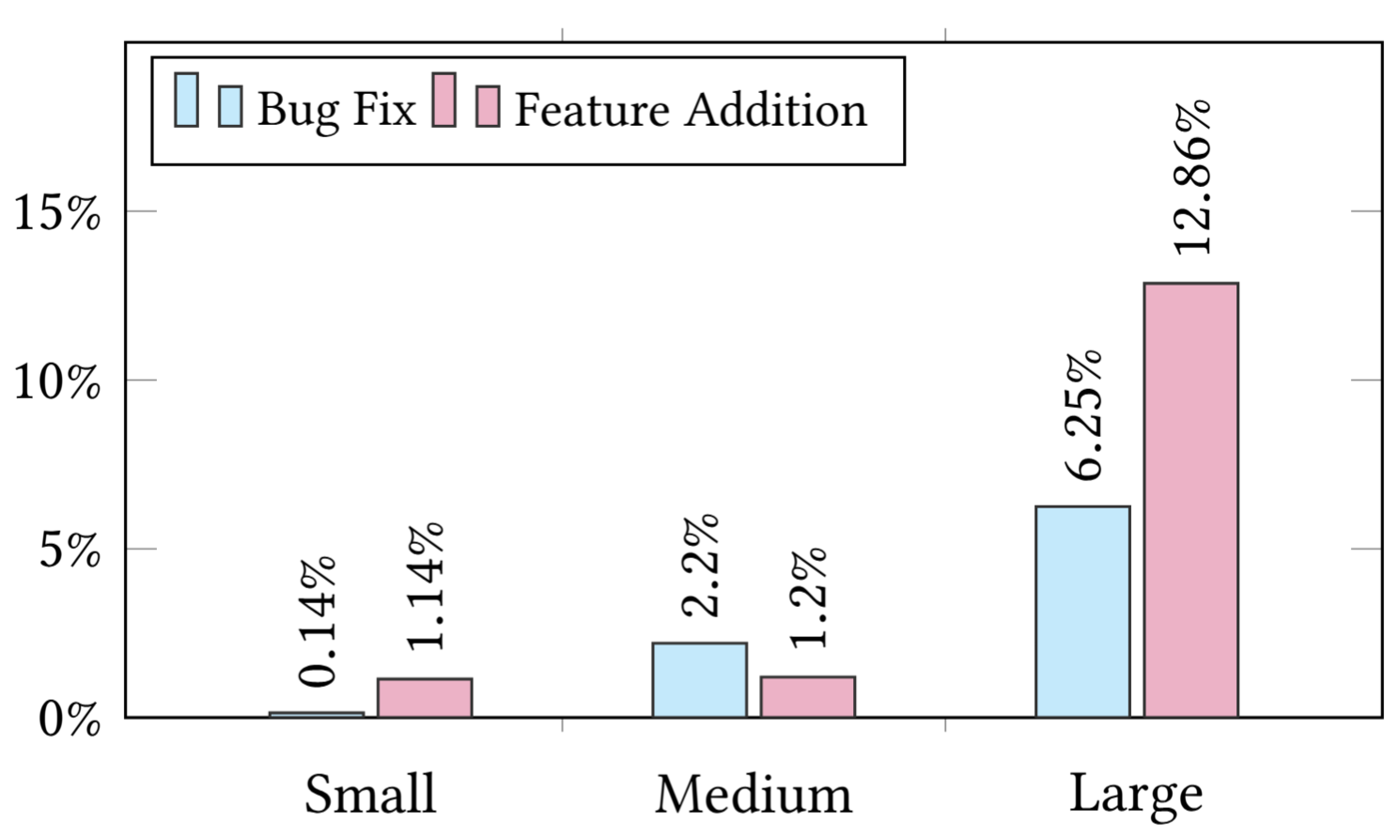}
    \vspace{-1em}
    \caption{Average of commit categories of female contributors}
\label{fig:commit_category}
\end{figure}

\noindent\textbf{Dynamic Impact of Female Contributions on AI Systems.} Borges et al. \cite{borges2016understanding} demonstrated that adding new features accelerates repository popularity, and our analysis (Fig.~\ref{fig:commit_category}) reveals that female contributors tend to prioritize feature addition over bug fixing. Given our findings from \textbf{RQ1} that diverse repositories are both more popular and enjoy greater community engagement, it is evident that female contributions play a significant role in enhancing the popularity of AI systems. However, as Trinkenreich et al. \cite{trinkenreich2022women} observed, women are more prevalent in community-centric roles, with nearly half of their contributions being non-code. These findings underscore the potential for further empowering women in AI systems by encouraging deeper involvement in the development and maintenance of AI systems.
% \vspace{-1em}

\begin{figure}
    \centering
    \includegraphics[width=0.9\linewidth]{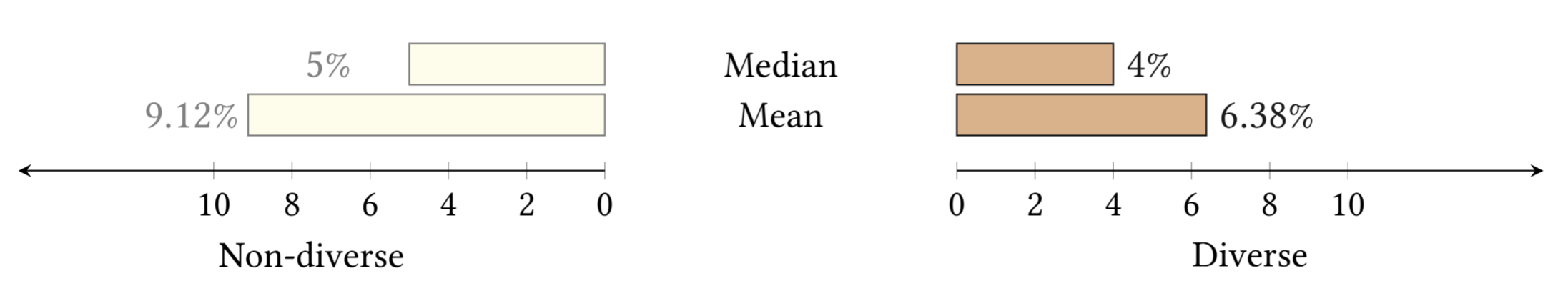}
    \caption{Mean and median value of SATD in diverse and non-diverse AI repositories. }
    \label{fig:SATD}
\end{figure}

\noindent\textbf{Mitigation of Technical Debt in Diverse Teams.} Our results from \textbf{RQ2} indicate that non-diverse repositories exhibit comparatively lower code quality than diverse ones.
Prior research has linked lower code quality with an increase in technical debt \cite{digkas2020can, lano2018technical}, including self-admitted technical debt \cite{bavota2016large, zampetti2017recommending}. Our analysis of 20 sample AI repositories validates these findings. As shown in Fig.~\ref{fig:SATD}, non-diverse teams contain more self-admitted technical debt (SATD) comments than diverse teams, with an average SATD occurrence approximately 30\% greater in non-diverse teams. This trend suggests that diverse teams, with their broader range of perspectives and more robust code review practices, not only produce more reliable code but are also more effective at minimizing and managing SATD. Overall, diverse teams appear to reduce and manage technical debt more efficiently.

\begin{figure}
    \centering
    \includegraphics[width=\linewidth]{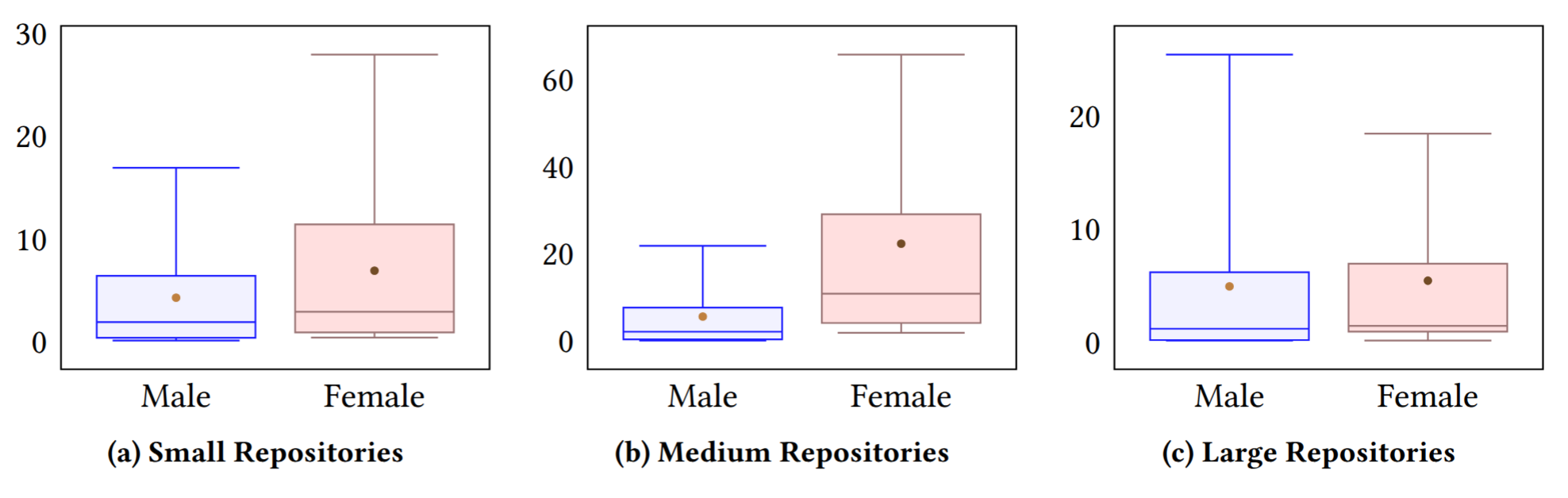}
    \caption{Contribution Ratio of male and female contributors across repositories ($\bullet$ presents the mean value).}
\label{fig:contributio_ratio}
\end{figure}

\begin{figure}
    \centering
    \includegraphics[width=\linewidth]{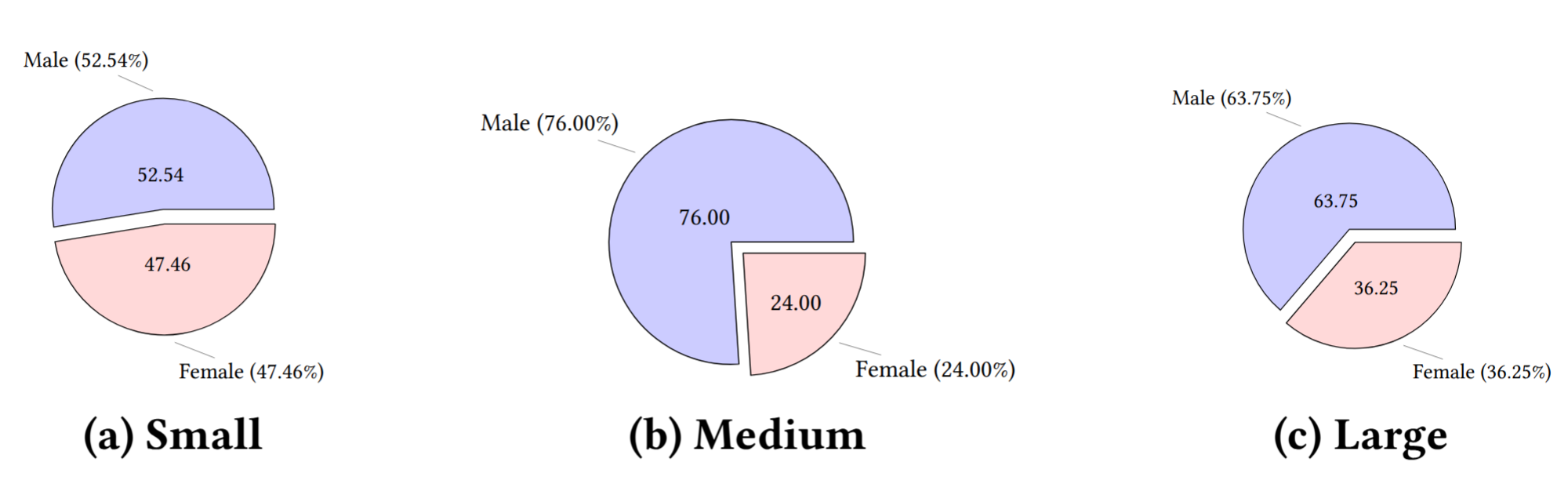}
    \caption{Average male and female contributor ratio}
    \label{fig:participation_ratio}
\end{figure}

\textbf{Quality in Scarcity: The Paradox of Female Contributions in the Development of AI Systems.}
We found in \textbf{RQ3} that lines of code written by female contributors demonstrate higher quality than male contributors. While we found that the lines of code written by female contributors are significantly low, Fig.~\ref{fig:contributio_ratio} shows that in terms of contribution ratio (calculated as number of contributions / total number of female contributors), female contributors are not far behind than male contributors. For example, in at least half of the small repositories, the median contribution ratio for female contributors is 3 compared to 2 for male contributors. Interestingly, in at least half of the medium repositories, the median contribution ratio for female contributors is 11, while the median contribution for male contributors is 2.25. However, the contribution ratio in the large repositories is almost the same for both male and female contributors. 
Moreover, as shown in Fig.~\ref{fig:participation_ratio}, the average ratio of female contributors remains disproportionately low, particularly in medium and large repositories, where female participation is half that of males. This disparity persists despite female contributors maintaining comparable contribution levels in large repositories and outperforming in smaller ones. Our findings align with previous research indicating that, despite their lower overall participation, female contributors tend to deliver higher contributions in the systems development. For instance, Terrell et al. \cite{terrell2017gender} demonstrated that women's pull requests are accepted at higher rates than men's, suggesting superior quality in their contributions. These findings underscore the need to address the underrepresentation of female contributors, as their elevated contribution quality is crucial for enhancing the reliability and sustainability of AI systems.

\section{Threats to Validity}
Threats to \textbf{external validity} concern the generalizability of our findings. Our analysis focuses on AI repositories associated with flagship AI conferences over a ten-year period. While this enables a systematic analysis of high-impact academic projects, it may not fully capture the development practices found in industry-led or community-driven open-source repositories.
Additionally, most repositories in our dataset are hosted on GitHub—the dominant platform for academic AI research. However, other platforms like GitLab or Bitbucket may adopt different collaboration practices, which could lead to varying outcomes. Similarly, our study primarily analyzes Python repositories, given Python’s prevalence in AI development. Projects written in other languages (e.g., C++, Java) may exhibit different patterns in terms of diversity and software quality.
Lastly, the field of AI continues to evolve rapidly, and development practices may shift over time. We encourage future studies to replicate and extend our work across broader ecosystems, languages, and hosting platforms. Our replication package is publicly available to support such efforts.

Threats to \textbf{internal validity} arise from experimental errors and biases. One key concern is the accuracy of gender classification. We relied on the genderComputer tool, which reports a high precision score (93\%); however, misclassifications may still occur, particularly for ambiguous names. Additionally, our binary approach to gender identification introduces limitations, as most existing automatic classifiers assume a binary gender model. Including non-binary and other gender identities would require direct input from contributors, which was beyond the scope of this study. Future research should adopt more inclusive methodologies to improve accuracy and representation.
Another potential limitation is the selection of repository engagement and popularity features. While our metrics were chosen based on prior studies \cite{fan2021makes}, alternative feature sets could provide different insights, potentially introducing bias in measuring repository popularity and quality.
Additionally, our study relies on SonarQube for evaluating code quality. While SonarQube is a widely used static analysis tool \cite{marcilio2019static, lenarduzzi2020sonarqube}, different tools may yield varying results, potentially influencing our conclusions. Another possible threat is that some repositories may already use SonarQube or similar tools in their CI/CD pipelines, which could confound our measurements. However, we conducted independent scans using a fresh SonarQube instance and found no explicit evidence of such tools in most README files or CI configurations. We acknowledge this limitation and encourage future work to explore its impact on reported quality metrics.

Threats to \textbf{conclusion validity} refer to the degree to which the conclusion drawn in this study is credible.
The study identifies correlations between gender diversity and repository characteristics (popularity, engagement, and code quality), however, these factors alone may not fully explain the observed outcomes. Other elements, such as team experience, project scope, and external contributions, may also play a significant role in influencing these results.

\vspace{-1em}
\section{Conclusion and Future Work}
This study highlights the significant impact of gender diversity on the characteristics of AI systems, particularly in terms of project popularity, code quality, and individual contributions. We analyzed 195 repositories, and our findings indicate that diverse AI repositories not only attract higher community engagement but also demonstrate superior code quality compared to non-diverse repositories. Furthermore, while female contributions remain underrepresented, their code demonstrates higher maintainability and lower complexity in collaborative settings as well. Our findings reinforce the value of diverse perspectives in AI development. and emphasize the need for more inclusive AI development practices to enhance software sustainability, mitigate biases, and improve overall system reliability. Given the experimental nature of AI systems, diversity may play a different role compared to non-AI software, which often relies on more rigid architectures \cite{shameer2023relationship}. Future work could explore this contrast by comparing diversity impacts across different software domains.
Additionally, our findings suggest actionable insights for stakeholders. For project maintainers, fostering inclusive collaboration environments can enhance both code quality and engagement. Educators and mentors should focus on removing barriers to women’s deeper involvement in AI system development. Finally, research funders and organizations can support gender-balanced teams through targeted incentives, as our findings highlight the tangible quality benefits of such diversity.

Future research can extend our findings by analyzing AI repositories across platforms like GitLab and Bitbucket and adopting a more inclusive gender classification model. Controlled experiments could further explore causal links between gender diversity and software quality. Additionally, examining socio-technical factors—such as mentorship, contribution patterns, and workplace policies—may help develop strategies to foster a more inclusive AI ecosystem.

\section{Acknowledgement}
This research is supported in part by the Natural Sciences and Engineering Research Council of Canada (NSERC) Discovery Grants program, the Canada Foundation for Innovation's John R. Evans Leaders Fund (CFI-JELF), and by the industry-stream NSERC CREATE in Software Analytics Research (SOAR).

\bibliographystyle{acm}
\bibliography{MANUSCRIPT}
\end{document}